\documentclass[preprint,aps]{revtex4}

\newcommand {\bea}{\begin{eqnarray}}
\newcommand {\eea}{\end{eqnarray}}
\newcommand {\be}{\begin{equation}}
\newcommand {\ee}{\end{equation}}

\usepackage{graphicx}

\begin{document}


\title{Many Body Methods and Effective Field Theory}

\author{T.~Sch\"afer, C.-W.~Kao and S.~R.~Cotanch}

\affiliation{Department of Physics, North Carolina State University,
Raleigh, NC 27695}

\begin{abstract}
In the framework of pionless nucleon-nucleon effective field
theory we study different approximation schemes for the nuclear 
many body problem. We consider, in particular, ladder diagrams 
constructed from particle-particle, hole-hole, and particle-hole 
pairs. We focus on the problem of finding a suitable starting point 
for perturbative calculations near the unitary limit $(k_Fa)\to\infty$ 
and $(k_F r)\to 0$, where $k_F$ is the Fermi momentum, $a$ is the 
scattering length and $r$ is the effective range. We try to clarify 
the relationship between different classes of diagrams and the large 
$g$ and large $D$ approximations, where $g$ is the fermion degeneracy 
and $D$ is the number of space-time dimensions. In the large $D$
limit we find that the energy per particle in the strongly interacting
system is 1/2 the result for free fermions.
\end{abstract}

\maketitle

\newpage

\section{Introduction}
\label{sec_intro}

  The nuclear many-body problem is of fundamental importance 
to nuclear physics \cite{Bethe:1971}. The traditional approach 
to the many-body problem is based on the assumption that nucleons 
can be treated as non-relativistic point-particles interacting 
mainly via two-body potentials. Three-body potentials, relativistic 
effects, and non-nucleonic degrees of freedom are assumed to give 
small corrections. The many-body Schr\"odinger equation is 
solved using a variety of methods that involve both variational  
and numerical aspects.

 Over the last several years an alternative approach to nuclear
physics based on effective field theory (EFT) methods has been 
applied successfully to the two and three-nucleon systems 
\cite{Weinberg:1990rz,Beane:2000fx,Bedaque:2002mn,Epelbaum:1999dj}. 
EFT methods have the advantage of being directly connected to QCD, 
and of providing a framework which is amendable to systematic 
improvements. The application of EFT to nuclear systems is 
complicated by the appearance of anomalously small energy scales. 
In the two-body system these small scales are reflected by the 
large neutron-neutron and neutron-proton scattering lengths, 
$a_{nn}(^1S_0)\simeq -18$ fm and $a_{np}(^3S_1)\simeq 5$ fm,
respectively. Effective field theories capable of describing 
systems with anomalously large scattering lengths require 
summing an infinite number of Feynman diagrams at leading 
order \cite{Weinberg:1990rz}.

 In this work we wish to study the EFT approach to the nuclear 
many body problem
\cite{Hammer:2000xg,Steele:2000qt,Kaiser:2001jx,Furnstahl:2002gt,Krippa:2002ht,Lee:2004si,Koch:1987me}.
We will focus on the equation of state of pure neutron matter at low 
to moderate density, a problem that is of relevance to the structure 
of neutron stars. The neutron matter problem has the theoretical 
advantage that there are no three-body forces at leading order. 
As in the two-body system the main obstacle is the large 
scattering length. If the scattering length was small the 
equation of state and other quantities of interest could 
be expanded in $(k_Fa)$, where $k_F$ is the Fermi momentum. 
This is the standard low density expansion for a hard sphere 
Fermi gas which was studied by Huang, Lee and Yang in the 1950's
and rederived in the EFT context by Hammer and Furnstahl
\cite{Huang:1957,Lee:1957,Hammer:2000xg}. In real nuclear 
matter, however, $|k_Fa|\gg 1$ and the perturbative low 
density expansion is not useful. 

 An interesting system that illustrates the difficulties of 
the nuclear matter problem is a dilute liquid of non-relativistic 
spin 1/2 fermions interacting via a short range potential with 
infinite scattering length. In this case the parameters that 
characterize the many body problem are either infinite or zero, 
$|k_Fa|\to\infty$ and $|k_Fr|\to 0$, where $a,r$ are the scattering 
length and the effective range. Dimensional analysis implies that
the equation of state is of the form 
\be
\label{xi}
\frac{E}{A}=\xi \frac{3}{5}\frac{k_F^2}{2m}, 
\ee
where $\xi$ is dimensionless number. For free fermions
$\xi=1$, but for strongly correlated fermions the theory contains 
no obvious expansion parameter and the determination of $\xi$ 
is a difficult non-perturbative problem. Recent interest in 
this problem has been fueled by experimental advances in 
creating cold, dilute gases of fermionic atoms tuned to be 
near a Feshbach resonance \cite{cold}. These experiments
are beginning to yield results for the equation of state 
of non-relativistic fermions in the limit $|k_Fa|\to\infty$. 

 A plausible strategy for investigating neutron matter is to
start from a numerical or variational solution of the 
``unitary limit'' system and to include corrections due to 
the finite effective range, explicit pion degrees of freedom, 
many body forces, etc.~perturbatively. Recent numerical studies 
of many body systems with a large scattering length can be found in 
\cite{Carlson:2003wm,Chen:2003vy,Wingate:2004wm,Lee:2004qd}.
In this work we study analytic many body approximations that 
could be used as the starting point for a theory of neutron matter 
based on EFT interactions. We focus on ladder diagrams built 
from particle-particle or particle-hole bubbles and study whether 
these approximations can be consistently renormalized and 
yield a stable $|k_Fa|\to\infty$ limit. We also examine the 
relationship of these approaches to the large $g$ and large 
$D$ limits, where $g$ is the number of fermion fields and $D$ 
is the number of space-time dimensions.

\section{Particle-particle ladder diagrams}
\label{sec_pp}

 We will consider non-relativistic fermions governed by an effective 
lagrangian of the form 
\be
\label{L_eff}
 {\cal L}_{\rm eff} =
 N^\dagger \left( i \partial_0 + \frac{\nabla^2}{2M}\right) N
   - \frac{C_0}{2} (N^\dagger N)^2 
   + \frac{C_2}{16}  \left[ (NN)^\dagger 
       (N{\stackrel{\leftrightarrow}{\nabla}}^2 N) +h.c. \right]   
  + \ldots \, ,
\ee
where $\stackrel{\leftrightarrow}{\nabla}=\stackrel{\leftarrow}{\nabla}
-\stackrel{\rightarrow}{\nabla}$. We have not displayed terms with 
higher derivatives or more powers of the fermion field, including 
two-derivative terms that act in the $p$-wave channel. The parameters 
$C_0$ and $C_2$ are related to the $s$-wave scattering length and the 
effective range. In the power divergence subtraction (PDS) scheme the 
relationship is given by \cite{Kaplan:1998we}
\be
\label{c0}
C_0 = - \frac{4\pi/M}{\mu-1/a}\ ,
\qquad
C_2 \, k^2 = \frac{4\pi/M}{(\mu-1/a)^2} \frac{r}2\; k^2 ,
\ee
where $\mu$ is the renormalization scale. The advantage of the 
PDS scheme is that the $C_i$ are of natural size even if the 
scattering length is large. Dimensional regularization with 
minimal subtraction gives $C_0=(4\pi a)/M$ which is unnatural 
in the limit $a\to\infty$. 

 We are interested in the energy density and pressure of a 
many body system with baryon density $\rho$. At zero temperature 
the density is related to the Fermi momentum via $\rho=gk_F^3/
(6\pi^2)$, where $g=2$ is the degeneracy factor. The free propagator 
is given by 
\be
\label{G_0}
G_0(k)_{\alpha\beta} = \delta_{\alpha\beta} 
\left( \frac{\theta(k-k_F)}{k_0-k^2/2M+i \epsilon}
+ \frac{\theta(k_F-k)}{k_0-k^2/2M-i \epsilon}
\right) \ ,
\ee 
and describes two types of excitations, holes with momentum $k<k_F$ 
and particles with $k>k_F$. Using this propagator and the vertices 
from equ.~(\ref{L_eff}) we can compute the energy per particle as 
a perturbative expansion in $(k_Fa)$. To order $(k_Fa)^2$ the result 
is \cite{Huang:1957,Lee:1957,Furnstahl:2002gt}
\be 
\label{E_pert}
\frac{E}{A} = \frac{k_F^2}{2M}\left[ 
 \frac{3}{5} + (g-1)\left(   \frac{2}{3\pi} (k_Fa) 
 + \frac{4}{35\pi^2} (11-2\log(2))(k_Fa)^2 \right)
 + O((k_Fa)^3) \right].
\ee
Effective range corrections appear at $O((k_Fa)^2(k_Fr))$ and if $g$ 
is bigger than 2 logarithmic terms appear at $O((k_Fa)^4\log(k_Fa))$. 

\begin{figure}
\includegraphics[width=6cm,clip=true]{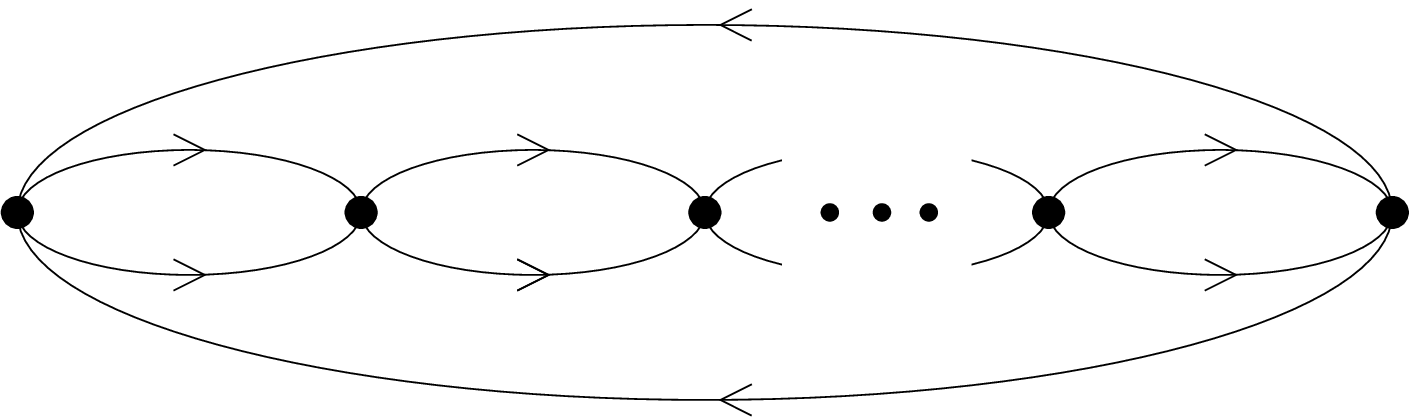}
\hspace*{1.5cm}
\includegraphics[width=6cm,clip=true]{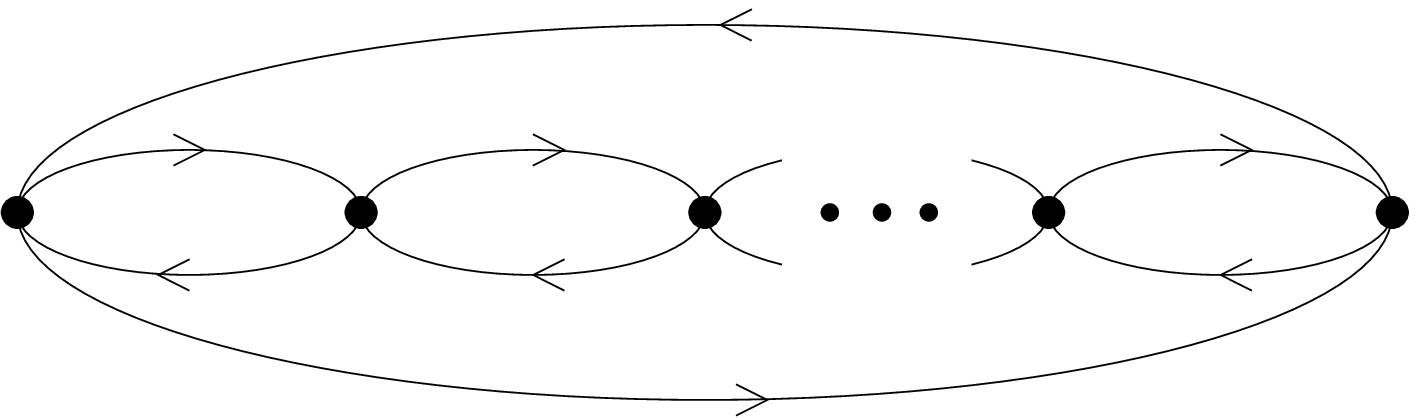}
\caption{Particle-particle ladder diagrams (left panel)
and particle-hole ring diagrams (right panel) in the
effective field theory. }
\label{fig_bub}
\end{figure}

 This expansion is clearly useless if $(k_Fa)\gg 1$. In the case 
of zero baryon density it is well known that an infinite set of 
bubble diagrams with the leading order contact interaction must 
be resummed if the two-body scattering length is large. It is 
natural to extend this calculation to non-zero baryon density by 
summing particle-particle bubbles with the finite density propagator 
given in equ.~(\ref{G_0}). In traditional nuclear physics this 
approach is known as Brueckner theory \cite{Fetter:1971,Day:1978}.
The elementary particle-particle bubble is given by 
\be
\label{fpp}
\int\!\! \frac{d^3q}{(2\pi)^3} \;\frac{\theta_q^+}{k^2-q^2+i\epsilon} 
 = - \frac{\mu}{4\pi} + \frac{k_F}{(2\pi)^2} f_{PP}(\kappa,s) \ .
\ee
We are following here the notation of Steele \cite{Steele:2000qt}.
The theta function $\theta_q^+\equiv\theta(k_1-k_F)\theta(k_2-k_F)$
with $\vec{k}_{1,2}=\vec{P}/2\pm \vec{k}$ requires both momenta
to be above the Fermi surface. The first term on the RHS is the 
vacuum contribution which contains the PDS renormalization scale 
$\mu$. The second term is the medium contribution which depends
on the scaled relative momentum $\vec{\kappa}=\vec{k}/k_F$ and 
center-of-mass momentum $\vec{s}=\vec{P}/(2k_F)$. For $s<1$ we have
\begin{figure}
\includegraphics[width=10cm,clip=true]{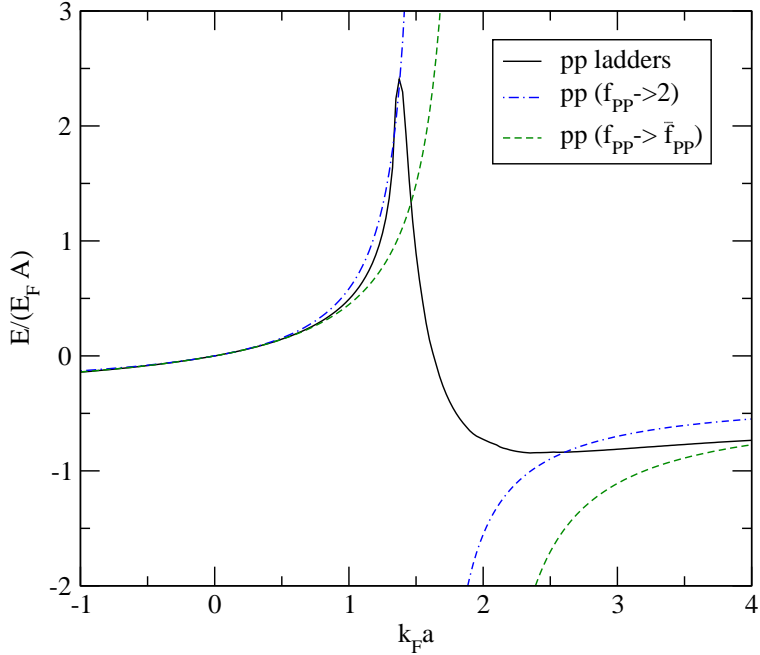}
\caption{Interaction energy per particle from particle-particle ladder 
diagrams as a function of $(k_Fa)$. The curves show a numerical 
calculation of the ladder sum and the two approximation $f_{PP}\to 2$ and 
$f_{PP}\to \langle f_{PP}\rangle$ discussed in the text.}
\label{fig_pp}
\end{figure}
\be
\label{fpp_2}
 f_{PP}(\kappa,s) = 1+s+
   \kappa \log \left|\frac{1+s-\kappa}{1+s+\kappa}\right|
   + \frac{1-\kappa^2-s^2}{2s} \log \left|
    \frac{(1+s)^2-\kappa^2}{1-\kappa^2-s^2}\right| \ .
\ee
Particle-particle ladder diagrams built from the elementary 
loop integral given in equ.~(\ref{fpp}) form a geometric 
series. The contribution of ladder diagrams to the energy 
per particle is given by \cite{Fetter:1971,Steele:2000qt}
\bea
\label{E_pp}
\frac{E_{PP}}{A} = \frac{3(g-1)\pi^2}{k_F^3}
\int\! \frac{d^3P}{(2\pi)^3}\frac{d^3k}{(2\pi)^3} 
 \; \theta_k^- \;
\frac{4\pi a/M}{1-\frac{k_Fa}{\pi} f_{PP}(\kappa,s)} \ .
\eea
This result can be interpreted as the trace of the in-medium 
particle-particle scattering matrix over all occupied (hole)
states. Note that equ.~(\ref{E_pp}) is independent of the renormalization
scale parameter $\mu$. This is in contrast to the perturbative
result equ.~(\ref{E_pert}) which is independent of $\mu$
only if $a$ is small. In general the integral in equ.~(\ref{E_pp})
has to be performed numerically. Steele suggested that in 
the large $D$ limit the function $f_{PP}$ can be replaced by its
asymptotic value 2 and we will examine this claim in Section \ref{sec_d}. 
Another possible approximation is to replace $f_{PP}$ by its 
phase space average
\be 
\langle f_{PP} \rangle = \frac{6}{35}
 \left( 11-2\log(2)\right),
\ee
where $\langle .\rangle$ denotes an average over all momenta
corresponding to occupied (hole) states. In this case we find
\be 
\frac{E_{PP}}{A} =(g-1)\frac{k_F^2}{2M} 
 \frac{2(k_Fa)/(3\pi)}{1-\frac{6}{35\pi}(11-2\log(2))(k_Fa)}.
\ee
This approximation has the virtue that the energy per 
particle from ladder diagrams agrees with the perturbative 
result up to $O((k_Fa)^2)$. In Fig.~\ref{fig_pp} we compare 
the two approximations with numerical results. We observe 
that all calculations agree fairly well if the scattering length
is either negative or positive and large. For $g=2$ the parameter 
$\xi$ in equ.~(\ref{xi}) is given by
\be
\xi=0.44\;\; (f_{PP}\to2),\hspace{0.5cm}
\xi=0.32\;\; (f_{PP}\to\langle f_{PP}\rangle ),\hspace{0.5cm}
\xi=0.24\;\; ({\rm num}).
\ee
For $g>2$ the results indicate that $\xi$ is negative and
the homogeneous low density phase is unstable. The different 
calculations shown in Fig.~\ref{fig_pp} disagree strongly in 
the regime $(k_Fa)\sim 1$. Approximating $f_{PP}$ by a 
constant leads to a singularity in the energy per particle. 
In the numerical calculation this singularity is smoothed
out, but a significant enhancement in the energy per particle
remains. However, even in this case the particle-particle
ladder sum has singularities for certain momenta that correspond
to occupied states. These singularities are presumably 
related to the existence of deeply bound two-body states
in the vacuum for $a\sim \mu^{-1}$. In this case interactions
between the bound states are essential and the approximations 
used in this section are not reliable.


\section{Effective range corrections and hole ladders}
\label{sec_range}

\begin{figure}
\includegraphics[width=10cm,clip=true]{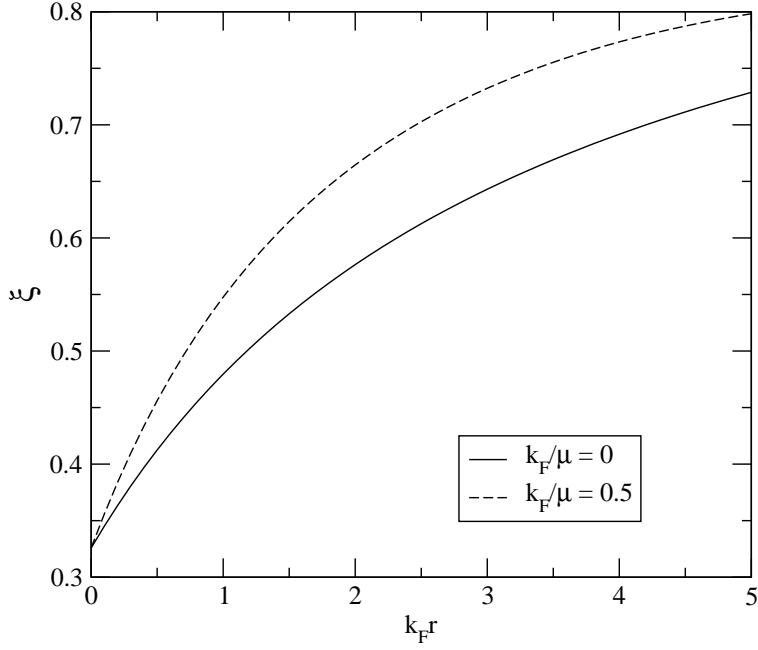}
\caption{
Equation of state of a dilute Fermi gas in the unitary limit 
$(k_Fa)\to\infty$ as a function of the effective range. We 
show the parameter $\xi$ defined in the text as a function 
of $(k_F r)$ for two different values of the Fermi momentum 
in units of the PDS renormalization scale $\mu$.}
\label{fig_eff_range}
\end{figure}

 In this section we study the question whether the ladder 
sum can be systematically improved by including higher 
order terms in the effective lagrangian. The two-derivative 
term proportional to $C_{2}$ incorporates effective range
corrections. At zero baryon density the particle-particle
scattering amplitude is \cite{Kaplan:1998we}
\be 
\label{T_2}
T(k) = \frac{C_0+C_2k^2}{1-\frac{M}{4\pi}(\mu+ik)
       (C_0+C_2k^2)}.
\ee
The effective range approximation corresponds to keeping 
$C_0$ to all orders but treating $C_2$ as a perturbation. 
The structure of equ.~(\ref{T_2}) is very simple because
in dimensional regularization, in both the MS and PDS 
renormalization schemes, powers of momentum internal to the 
elementary particle-particle loop diagrams are converted 
into powers of the external momentum. At finite density 
the integrals are more complicated. We discuss the 
calculation of the bubble sum in the appendix and only 
present the results here. The energy per particle is
\bea
\label{E_pp_C2}
\frac{E_{PP}}{A} &=& \frac{3(g-1)\pi^2}{k_F^3}
\int\! \frac{d^3P}{(2\pi)^3}\frac{d^3k}{(2\pi)^3} 
 \; \theta_k^- \frac{1}{1-MC_{0}I_{0}}  \nonumber \\
& & \hspace{0.3cm} \times  \Bigg\{ C_0 +
  C_{2}\frac{(k^2+MC_{0}g_{2})(4-MC_{2}g_{2})+MC_{2}g_{4}}
      {(2-MC_{2}g_{2})^2-(4C_{0}+4C_{2}k^2-C_{2}^{2}
   M(k^2g_{2}-g_{4}))MI_{0}} \Bigg\},
\eea
where $I_{n}$ and $g_{n}$ are 
\bea
\label{fpp_qn}
 I_{n}(k_F,\kappa,s) &=& 
   \int\!\! \frac{d^3q}{(2\pi)^3} \;
   \frac{q^{n}\theta_q^+}{k^2-q^2+i\epsilon} , \\
-g_{n}(k_F,s)&=& \int\!\! \frac{d^3q}{(2\pi)^3} 
   \;q^{n-2} \theta_q^{+}  ,
\eea
and satisfy the following relations
\bea
I_{0}(k_F,\kappa,s) &=& - \frac{\mu}{4\pi} 
  + \frac{k_F}{(2\pi)^2} f_{PP}(\kappa,s), \\
I_{2}(k_F,\kappa,s) &=& k^2 I_{0}(k_F,\kappa,s)+g_{2}(k_{F},s),\\
I_{4}(k_F,\kappa,s) &=& k^2 I_{2}(k_F,\kappa,s)+g_{4}(k_{F},s).
\eea
The explicit forms of $g_{2}$ and $g_{4}$ are
\bea
g_{2}&\equiv & \frac{k_{F}^{3}}{\pi^2}\bar{g}_{2}(s) 
  =\frac{k_{F}^{3}}{\pi^2}\left[
   -\frac{1}{3}+\theta(1-s)\left(
   \frac{1}{6}-\frac{s}{4}+\frac{s^3}{12}\right)\right], \\
g_{4}&\equiv &\frac{k_{F}^{5}}{\pi^2}\bar{g}_{4}(s) =
 \frac{k_{F}^{5}}{\pi^2}\left[-\frac{1}{5}-\frac{s^2}{3}
 +\theta(1-s)(1-s)^{3}\left(
 \frac{1}{10}+\frac{s}{20}+\frac{s^2}{60}\right)\right],
\eea
and both $g_{2}$ and $g_{4}$ vanish as $k_F\to 0$. If $k_F^2C_{2}\ll C_{0}$ 
then equ.~(\ref{E_pp_C2}) can be simplified to
\bea
\label{E_pp_C2_Pert}
\frac{E_{PP}}{A} = \frac{3(g-1)\pi^2}{k_F^3}
  \int\! \frac{d^3P}{(2\pi)^3}\frac{d^3k}{(2\pi)^3} 
   \; \theta_k^- \frac{1}{1-C_{0}MI_{0}}\left\{C_{0}
    + C_{2}\frac{k^2+MC_{0}g_{2}}{1-C_{0}MI_{0}}
\right\}
\eea
Using equ.~(\ref{c0}) to relate the coupling constants $C_0$ and 
$C_2$ to the scattering length and the effective 
range we find 
\bea
\label{E_pp_C2_Pert_PDS}
\frac{E_{PP}}{A} &=& \frac{3(g-1)\pi^2}{k_F^3}
    \int\! \frac{d^3P}{(2\pi)^3}\frac{d^3k}{(2\pi)^3} 
    \; \theta_k^- \;
    \frac{4\pi a/M}{1-\frac{k_Fa}{\pi} f_{PP}(\kappa,s)}
    \nonumber \\
 && \;\times \left\{1+\frac{(k_{F}a)(k_{F}r)}{1-\frac{k_Fa}{\pi} 
   f_{PP}(\kappa,s)}
  \left[\frac{\kappa^2}{2}+\frac{k_{F}a}
  {(k_{F}a)(\mu/k_{F})-1}\cdot\frac{2\bar{g}_{2}}{\pi}\right]\right\}.
\eea
We observe that the energy per particle depends on the 
renormalization scale $\mu$. We are particularly interested in the
situation when  $|k_Fa|\gg 1$ and $|k_Fr|< 1$ for which
\bea
\frac{E_{PP}}{A} &=& \frac{3(g-1)\pi^2}{k_F^3}
\int\! \frac{d^3P}{(2\pi)^3}\frac{d^3k}{(2\pi)^3} 
 \; \theta_k^- \;
\frac{4\pi a/M}{1-\frac{k_Fa}{\pi} f_{PP}(\kappa,s)}\nonumber \\
&& \times \left\{1+\frac{(k_{F}a)(k_{F}r)}{1-\frac{k_Fa}{\pi} 
  f_{PP}(\kappa,s)}\left[\frac{\kappa^2}{2}+\frac{k_{F}}
  {\mu}\cdot\frac{2\bar{g}_{2}}{\pi}\right]\right\}.
\eea
We find that effective range corrections are small and
independent of $\mu$ provided $|k_Fr|<1$ and $k_F/\mu<1$.
Evaluating the integral by replacing $f_{PP}$ and $g_2$ 
by their phase space averages and taking $k_Fa\to \infty$
we get
\be 
\xi (k_Fr) = 0.32 + 0.19 (k_Fr) + O(k_F/\mu,(k_Fr)^2).
\ee
Using equ.~(\ref{E_pp_C2}) we can also study the behavior
of the universal parameter $\xi$ for larger values of 
$k_Fr$. The result is shown in Fig.~\ref{fig_eff_range}.
We observe that the dependence of $\xi$ on $k_Fr$ becomes
weaker as $k_Fr$ grows. In the limit $k_Fr\to\infty$ 
the parameter $\xi$ slowly approaches the free Fermi 
gas value $\xi=1$ \cite{Schwenk:2005ka}.

\begin{figure}
\includegraphics[width=10cm,clip=true]{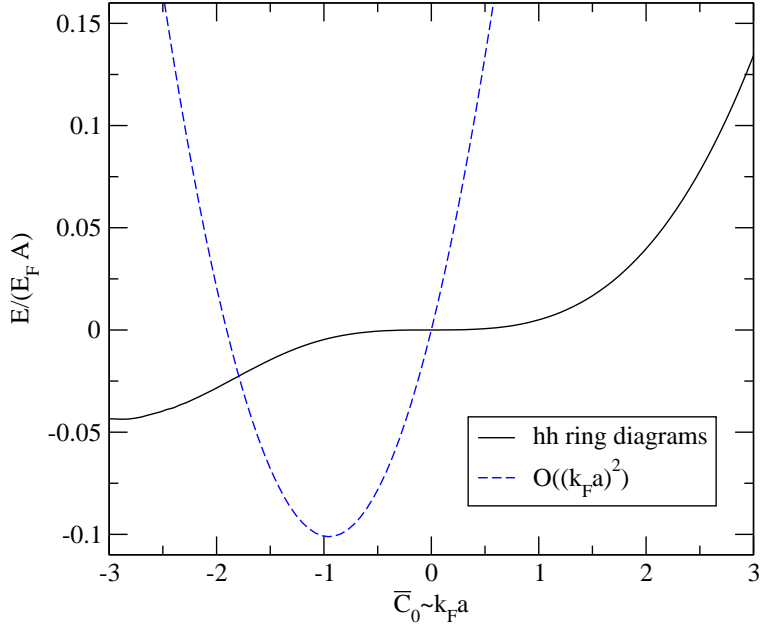}
\caption{
Energy per particle from hole-hole ladder diagrams as a 
function of $\bar{C}_0=Mk_FC_0/(4\pi)$. $\bar{C}_0$ is 
equal to $(k_Fa)$ in the $\overline{MS}$ scheme, but in the 
PDS scheme the relation between $\bar{C}_0$ and $(k_Fa)$ 
depends on the renormalization scale $\mu$. We also show 
the perturbative result up to order $\bar{C}_0^2$.}
\label{fig_hh}
\end{figure}

 In addition to higher order corrections to the effective 
interaction we also consider larger classes of diagrams. 
A simple extension of the calculation of the particle-particle
ladder sum is the inclusion of hole-hole ladders. Following the 
steps that lead to equ.~(\ref{E_pp}) we find
\bea
\label{E_hh}
\frac{E_{HH}}{A} = \frac{3(g-1)\pi^2}{k_F^3}
\int\! \frac{d^3P}{(2\pi)^3}\frac{d^3k}{(2\pi)^3} 
 \; \theta_k^+ \left\{
\frac{C_0}{1-\frac{k_FMC_0}{4\pi^2} f_{HH}(\kappa,s)} 
 - C_0 -\frac{k_FMC_0^2}{4\pi^2} f_{HH}(\kappa,s)
 \right\} ,
\eea
where $f_{HH}(\kappa,s)=f_{PP}(\kappa,-s)$ is the hole-hole
bubble. We have subtracted the first two terms in the expansion of the 
geometric series. These terms have UV divergences and need to be 
treated separately. In our case this is not necessary since the
two contributions are already included in the particle-particle 
ladder sum. We observe that the remaining part of the hole ladders
is finite and only depends on $C_0$ and not the PDS renormalization
scale $\mu$. This implies that if the coupling constant is related to 
the scattering length according to equ.~(\ref{c0}), the energy per 
particle will depend on the renormalization scale $\mu$. Numerical 
results for the hole-ladders are shown in Fig.~\ref{fig_hh}. We 
observe that if $C_0$ is of natural size the energy per particle 
from hole ladders is indeed very small compared to the contribution
from particle ladders. 

\section{Particle-hole ring diagrams and the large $g$ expansion}
\label{sec_ph}

 Another important class of diagrams is the set of particle-hole 
ring diagrams. In gauge theories ring diagrams play a crucial 
role since they incorporate screening corrections and their 
inclusion of is necessary to achieve a well behaved perturbative 
expansion. In theories with short range 
interactions particle-hole bubbles also provide important corrections 
to the effective interaction. For example, particle-hole screening 
corrections reduce the s-wave BCS gap by a factor $\sim 1/2$ in the 
weak coupling limit.

 The real and imaginary parts of the particle-hole bubble are
given by \cite{Fetter:1971}
\bea
\label{pi_ph}
{\rm Re}\, \Pi_0(\nu,q)&=&\frac{Mk_F}{4\pi^2}
\Bigg\{-1+\frac{1}{2q}
 \bigg(
   1 - \bigg(\frac{\nu}{q} - \frac{q}{2}\bigg)^2
 \bigg)
 \ln\left|\frac{1+ (\nu/q-q/2)}{1-(\nu/q-q/2)}\right| \nonumber\\
&&\mbox{}\hspace{2.0cm} -\frac{1}{2q}
 \bigg(
   1 - \bigg(\frac{\nu}{q} + \frac{q}{2}\bigg)^2
 \bigg)
 \ln\left|\frac{1+ (\nu/q+q/2)}{1-(\nu/q+q/2)}\right|\Bigg\}
\label{eq:repi} \\
{\rm Im}\, \Pi_0(\nu,q) &=&-\frac{Mk_F}{8\pi q} 
\left\{
\begin{array}{ccc}
 1-\left(\frac{\nu}{q} - \frac{q}{2}\right)^2 
  &  q>2,\; & \frac{q^2}{2}+q \geq \nu \geq \frac{q^2}{2}-q\,, \\
 1-\left(\frac{\nu}{q} - \frac{q}{2}\right)^2
  &  q<2,\; & q+\frac{q^2}{2} \geq \nu \geq q-\frac{q^2}{2}\,, \\
 2\nu
  &  q<2,\; & 0 \leq \nu \leq q-\frac{q^2}{2} \, ,
\end{array}\right.
\eea
where $\nu= k_0M/k_F^2$ and $q=|\vec{k}|/k_F$. Ring diagrams 
containing particle-hole bubbles can be summed in essentially the 
same way as the particle-particle ladders. The main difference 
arises from different spin and symmetry factors. The spin factor 
of the $n$-th order particle-particle ladder contribution is 
$g(g-1)2^n$. The spin factors of particle-hole diagrams are 
more complicated. The situation simplifies in the limit of large 
$g$, often called the large $N$ limit, as it is equivalent to 
the limit of a large number $N$ of degenerate spin 1/2 fermions. 
In this case every particle-hole bubble contributes a factor $g$.
Indeed, one can show that the particle-hole ring diagrams are 
the leading diagrams in the large $g$ limit \cite{Furnstahl:2002gt}.

\begin{figure}
\includegraphics[width=10cm,clip=true]{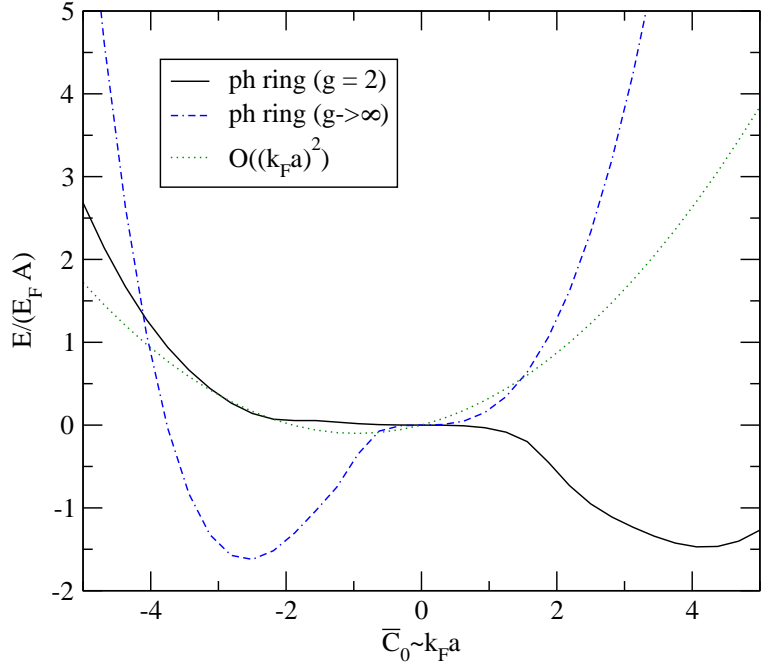}
\caption{Energy per particle from particle-hole ring diagrams
as a function of $\bar{C}_0=Mk_FC_0/(4\pi)$. $\bar{C}_0$ is 
equal to $(k_Fa)$ in the $\overline{MS}$ scheme, but in the 
PDS scheme the relation between $\bar{C}_0$ and $(k_Fa)$ 
depends on the renormalization scale $\mu$. We also show the 
perturbative result up to order $\bar{C}_0^2$.}
\label{fig_ph}
\end{figure}

 The $n$-th order diagrams in both the particle-particle ladder 
sum and the particle-hole ring sum have symmetry factors $1/n$.
In the case of the ladder diagrams this factor is canceled by a 
factor $n$ specifying the $n$ different ways in which the diagram 
can be cut to represent it as a particle-particle Green function 
integrated over all occupied states. For the ring diagrams it is 
more convenient to carry out the energy integration explicitly, 
and no factor $n$ appears. As a consequence, the sum of all ring 
diagrams is a logarithm. In the large $g$ limit we find
\be 
\label{E_ring}
 E = -\frac{i}{2} \int \frac{d^4k}{(2\pi)^4} 
  \left[ \log\left(1-gC_0\Pi(k)\right) + gC_0\Pi_0(k)
    + \frac{1}{2} \left(gC_0\Pi_0(k)\right)^2 \right],
\ee
where we have subtracted the first two terms in the expansion 
of the logarithm in order to make the integral convergent. These
two terms can be computed separately and correspond to the first 
two terms of the perturbative expansion given in equ.~(\ref{E_pert}).
Equ.~(\ref{E_ring}) shows that the correct way to take the large 
$g$ limit is to keep $gC_0$ constant as $g\to\infty$. In this 
case the free Fermi gas contribution as well as the $O(k_Fa)$
correction in equ.~(\ref{E_pert}) is of order $O(1)$ while the
ring diagrams give a correction of order $O(1/g)$.

Since the energy is real, the integral in equ.~(\ref{E_ring}) 
can be written as \cite{Fetter:1971,Furnstahl:2002gt}
\bea
\label{E_ring_2}
\frac{E}{A} &=& \frac{3}{g\pi}\left(\frac{k_F^2}{2M}\right)
 \int_0^\infty  q^2 dq  \int_0^\infty d\nu \Bigg[
  gC_0\, {\rm Im}\, \Pi_0(\nu,q)+(gC_0)^2\, {\rm Im}\, \Pi_0(\nu,q)
{\rm Re}\, \Pi_0(\nu,q) \nonumber\\
& & \hspace{1cm}\mbox{}     -\arctan \left(
  \frac{gC_0\, {\rm Im}\,\Pi_0(\nu,q)}
       {gC_0{\rm Re}\,\Pi_0(\nu,q)-1}\right) \Bigg].
\eea
In the PDS scheme $C_0$ is related to the scattering length by 
equ.~(\ref{c0}) and the ring energy depends of the renormalization 
scale. Numerical results (for $g=2$) are shown in Fig.~\ref{fig_ph}.
For simplicity we have taken $\mu=0$. In this case the energy 
per particle goes to infinity as $(k_Fa)\to\infty$. For other
values of $\mu$ the energy is finite, but strongly dependent 
on $\mu$. We also observe that the ring energy per particle is
less than $3E_F/5$ for $(k_Fa)<0.9$, which implies an instability
of the homogeneous system. 
 
 We emphasized above that equ.~(\ref{E_ring}) is correct 
only in the large $g$ limit. We can also compute the ring sum 
for $g=2$. In this case there are two possible channels with 
total spin zero and one. The $n$-th order ring diagram has 
spin factor $(-2,4,-2,4,\ldots)$. The sum of all ring diagrams
is
\bea 
\label{E_ring_3}
 E &=& -\frac{i}{2} \int \frac{d^4k}{(2\pi)^4} 
  \Big[ \log\left(1-C_0\Pi(k)\right) 
  +3\log\left(1+C_0\Pi(k)\right)   \nonumber \\
 & & \hspace{5cm}\mbox{}  
    - 2 C_0\Pi_0(k)
    + 2 \left(C_0\Pi_0(k)\right)^2 \Big].
\eea
The integral can be calculated as in equ.~(\ref{E_ring_2}) and
the result is shown in Fig.~\ref{fig_ph}. We observe that although
the correct $g=2$ result is quite different from the $g\to\infty$
result evaluated at $g=2$, qualitative features, like $E/A\to\infty$ 
as $C_0\to\infty$ and the presence of an unstable regime with
$E/A<3E_F/5$, remain unchanged.

\section{Large $D$ expansion}
\label{sec_d}

 In the previous section  we saw that the particle-hole ring energy
can be interpreted as the leading order contribution to the
energy in the large $g$ limit. This raises the question whether 
there exists an expansion that gives the ladder sum as the leading 
order contribution. In a very interesting paper Steele suggested
that expansing in $1/D$, where $D$ is the number of space-time 
dimensions, is the desired scheme \cite{Steele:2000qt}. If true 
the $1/D$ expansion offers a systematic approach to the fermion 
many-body system in the limit $(k_Fa)\to\infty$.

 The main idea is that many body diagrams in a degenerate 
Fermi system are very sensitive to the available phase space
and that the scaling behavior of phase space factors in the 
large $D$ limit could be a basis for a geometric expansion.  
Steele argued that the $1/D$ expansion corresponds to the 
hole-line expansion in traditional nuclear physics, and that 
it is consistent with EFT power counting for systems with 
a large scattering length. In this section we shall examine 
these claims in more detail. 

\begin{figure}
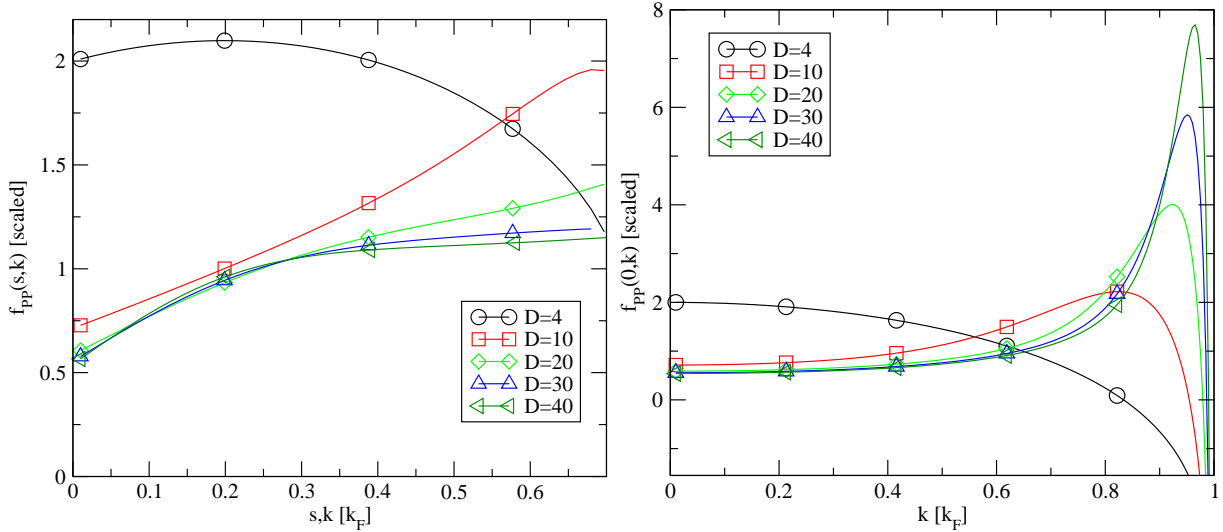

\includegraphics[width=8cm,clip=true]{fpp_d_scal.eps}
\includegraphics[width=8cm,clip=true]{bcs_d_scal.eps}
\caption{Scaled particle-particle scattering amplitude $Df_{PP}^{(D)}
(\kappa,s)$ for $\kappa=s$ (left panel) and $s=0$ (right panel). We 
observe that except in the BCS limit $s=0,\kappa\to 1$ the function 
$Df_{PP}^{(D)}$ approaches a smooth limit as $D\to\infty$. }
\label{fig_fppd}
\end{figure}

 The spatial density of a free Fermi gas in $D$ space-time dimensions 
is given by 
\be 
\label{rho_d}
 \rho = \frac{\Omega_{D-1}}{(2\pi)^{D-1}}
   \frac{k_F^{D-1}}{D-1}, 
  \hspace{0.5cm} 
  \Omega_{D-1}=\frac{2\pi^{(D-1)/2}}{\Gamma(\frac{D-1}{2})},
\ee
where $\Omega_D$ is the surface area of a $D$-dimensional unit 
ball. In the following we will always take the degeneracy factor 
$g=2$. The energy per particle is 
\be
\frac{E_0}{A} = \frac{D-1}{D+1}\left(\frac{k_F^2}{2M}\right)
  = \left\{ 1 - \frac{2}{D}+\ldots\right\}
  \left(\frac{k_F^2}{2M}\right).
\ee
We can compute perturbative corrections to this result
in $D$ dimensions. To leading order in $C_0$ we find
\be 
\frac{E_1}{A} = \frac{1}{D-1}  \left[
  \frac{\Omega_{D-1}C_0k_F^{D-3}M}{(2\pi)^{D-1}} \right]
 \left(\frac{k_F^2}{2M}\right) .
\ee
This expression indicates that in the weak coupling limit
the large $D$ limit should be taken in such a way that 
\be
\label{d_scal}
 \lambda \equiv  \left[ 
  \frac{\Omega_{D-1}C_0k_F^{D-3}M}{D (2\pi)^{D-1}} \right]
 \stackrel{D\to\infty}{\longrightarrow} {\it const} .
\ee
In the following we wish to study whether this limit
is smooth even if the theory is non-perturbative, and 
what class of diagrams is dominant. For this purpose
we consider the in medium particle-particle bubble for 
an arbitrary number of space-time dimensions $D$. The
result is 
\be 
\int\!\! \frac{d^{D-1}q}{(2\pi)^{D-1}} \;
  \frac{\theta_q^+}{k^2-q^2+i\epsilon} 
 = f_{vac}(k) + \frac{k_F^{D-3}\Omega_{D-1}}{2(2\pi)^{D-1}} 
   f_{PP}(\kappa,s),
\ee
where 
\be 
\label{fpp_d}
f_{PP}^{(D)}(\kappa,s)  = 
  \frac{2}{c_D}\int_{\sqrt{1-s^2}}^{1+s} dt\,
  c\left(D,x_0(s,t)\right)
  \frac{t^{D-2}}{\kappa^2-t^2}
  - 2I_2(D,s,\kappa)
\ee
with $x_0(s,t)=(s^2+t^2-1)/(2st)$. The factor 2 was inserted
in equ.~(\ref{fpp_d}) so that the normalization of $f_{PP}^{(D)}$ 
is consistent with the previous case $D=4$. The other terms are
\bea
 c(D,x_0) &=& 2x_0 \;\mbox{}_2F_1
   \left(\frac{1}{2},2-\frac{D}{2},\frac{3}{2},x_0^2\right), \\
  c_D\equiv c(D,1)&=&\sqrt{\pi}
   \frac{\Gamma\left(\frac{D}{2}-1\right)}
        {\Gamma\left(\frac{D-1}{2}\right)}, \\
 I_2(D,s,\kappa) &=& \frac{\mbox{}_2F_1\left(
           1,\frac{D-1}{2},\frac{D+1}{2},\frac{1+s}{\kappa^2}\right)}
        {(D-1)\kappa^2(1+s)^{1-D}} ,
\eea
where $\mbox{}_2F_1(a,b,c,z)$ is the hypergeometric function.
Numerical results for the function $f_{PP}^{(D)}(s,\kappa)$ are 
shown in Fig.~\ref{fig_fppd}. We observe that the particle-particle
bubble scales as
\be 
f_{PP}^{(D)}(s,\kappa) = \frac{1}{D} f_{PP}^0(s,\kappa)
 \left(1+ O\left(\frac{1}{D}\right)\right). 
\ee
The scaling behavior can be verified analytically in certain limits.
We find, in particular, 
\be 
f_{PP}^{(D)}(0,0) = \frac{2}{D-3}.
\ee
There is a subtlety associated with the BCS singularity at 
$s=0,\kappa\to 1$. Fig.~\ref{fig_fppd}b shows that the logarithmic
singularity is not suppressed by $1/D$. We observe, however, 
that the range of momenta for which $f_{PP}$ is enhanced
shrinks to zero as $D\to\infty$. We will study pairing 
in the large $D$ limit in Sect.~\ref{sec_bcs}.

Since $f_{PP}^{(D)}\sim 1/D$ we conclude that if the large $D$ 
limit is taken according to equ.~(\ref{d_scal}) then all ladder 
diagrams with particle-particle bubbles are of the same order 
in $1/D$. The sum of all ladder diagrams can be calculated by
noting that, except for the logarithmic (BCS) singularity at 
$s=0,\kappa=1$, the particle-particle bubble is a smooth function
of the kinematic variables $s$ and $\kappa$. Hole-hole phase space, 
on the other hand, is strongly peaked at $\bar{s}=\bar{\kappa}=
1/\sqrt{2}$ in the large $D$ limit. This can be seen by re-expressing
$s,\kappa$ in terms of $k_{1,2}$ and using $\bar{k}_{1,2}\to k_F$ 
as $D\to\infty$. If the phase space is strongly peaked we can replace 
the function $f_{PP}^{(D)}(s,\kappa)$ by its value at $\bar{s},
\bar\kappa$. We have not been able to calculate  the large $D$ limit 
of $f_{PP}^{(D)}(\bar{s},\bar{\kappa})$ analytically. Our numerical 
results show that $\lim_{D\to\infty} f_{PP}^{(D)}(\bar{s},\bar{\kappa})
/f_{PP}^{(D)}(0,0)=2.02\pm 0.02$. We therefore conjecture that 
$f_{PP}^{(D)}(\bar{s},\bar{\kappa})=4/D\cdot (1+O(1/D)))$.

 We illustrate the method by calculating the second order 
correction. This contribution involves an integral of the 
particle-particle bubble over hole-hole phase space. We 
find
\be 
\int\frac{d^{D-1}P}{(2\pi)^{D-1}}
 \int\frac{d^{D-1}k}{(2\pi)^{D-1}}\theta_k^-
 f_{PP}^{(D)}(\kappa,s)
 = \frac{k_F^{2D-2}}{D^2}
   \left[\frac{\Omega_{D-1}}{(2\pi)^{D-1}}\right]^2
  \frac{4}{D}\left(1+O\left(\frac{1}{D}\right)\right),
\ee
and the energy per particle is given by
\be
\frac{E_2}{A} = 2  \left[
  \frac{\Omega_{D-1}C_0k_F^{D-3}M}{D(2\pi)^{D-1}} \right]^2
 \left(\frac{k_F^2}{2M}\right).
\ee
Higher order terms can be calculated in the same fashion. 
As in $D=4$, particle-particle ladder diagrams sum to 
a geometric series. We get 
\be
\label{E_pp_d}
 \frac{E}{A} = \left\{ 1 + \frac{\lambda}{1-2\lambda} +
 O\left(\frac{1}{D}\right)\right\}
\left(\frac{k_F^2}{2M}\right),
\ee
where $\lambda$ is the coupling constant defined in equ.~(\ref{d_scal}).
We observe that if the strong coupling limit $\lambda \to \infty$ is taken
after the limit $D\to\infty$ the universal parameter $\xi$ is given 
by 1/2. 

 We have been able to compute the particle ladder contribution
to the energy per particle in the large $D$ limit. Steele argued
that all other contributions are suppressed by powers of $1/D$
since each additional hole line involves an integral over the Fermi 
surface of the type shown in equ.~(\ref{rho_d}) which gives a least 
one power of $1/D$. It is not clear if this argument is entirely 
correct. We have found, for example, that the main contribution of 
the particle-particle bubble also scales as $1/D$. We study this 
problem in more detail in Secs.~\ref{sec_bcs} and \ref{sec_scr}
where the potentially relevant pairing and screening corrections 
are examined, respectively.

\section{Pairing in the large $D$ limit}
\label{sec_bcs}

 In the previous section we noticed that in the large $D$ 
limit the particle-particle bubble is enhanced when $s=0$ and 
$\kappa\to 1$. In this limit the two particles are on opposite 
sides of the Fermi surface and the logarithmic enhancement of 
$f_{PP}$ is the well known BCS singularity. The result suggests
that in the large $D$ limit the pairing energy might dominate 
all other contributions to the energy density. In this section
we shall study this question by computing the BCS gap and the 
pairing energy in the large $D$ limit. 

 If the interaction is weak and attractive we can derive the standard 
BCS gap equation (see, for example, reference \cite{Schafer:2003vz})
\be
\label{gap}
\Delta = \frac{|C_0|}{2}\int \frac{d^{D-1}p}{(2\pi)^{D-1}}
  \frac{\Delta}{\sqrt{\epsilon_p^2+\Delta^2}}
\ee
with $\epsilon_p=E_p-E_F$ and $E_p=p^2/(2M)$. The integral in 
equ.~(\ref{gap}) can be carried analytically for arbitrary $D$.
We find \cite{Papenbrock:1998wb,Marini:1998}
\be
\label{gap2}
 1 = \frac{D\lambda\pi}{\sin(\pi\alpha)}
 \left(1+x^2\right)^{\alpha/2}
 P_\alpha \left( - \frac{1}{\sqrt{1+x^2}}\right)
\ee
where $\lambda$ is the dimensionless coupling constant defined in 
equ.~(\ref{d_scal}), $x =\Delta/E_F$ is the dimensionless gap, 
$P_\alpha(z)$ is the Legendre function and $\alpha=(D-3)/2$. We 
note that by going to arbitrary $D$ we have regularized the UV 
divergence in the gap equation using dimensional regularization. 
If the gap is small, $x\ll 1$, equ.~(\ref{gap2}) can be solved 
using the asymptotic behavior of the Legendre function $P_\alpha(z)$ 
near the logarithmic singularity at $z=-1$ \cite{Erdelyi:1953}
\be
P_\alpha(z) \sim \frac{\sin(\alpha \pi)}{\pi}
 \left( \log\left(\frac{1+z}{2}\right) + 
 2\gamma+2\psi(\alpha+1) + \pi\cot(\alpha\pi) \right) .
\ee
To leading order in $1/D$ we can also use the asymptotic
expression for the Digamma function $\psi(\alpha)=\Gamma'(\alpha)
/\Gamma(\alpha)\simeq \log(\alpha)+O(1/\alpha)$. We find 
\be 
\Delta = \frac{2e^{-\gamma}E_F}{D} 
  \exp\left(-\frac{1}{D\lambda}\right)
  \left(1+O\left(\frac{1}{D}\right)\right),
\ee
where $\gamma\simeq 0.5772$ is Euler's constant. We observe that 
the exponential suppression of the gap disappears if the large 
$D$ limit is taken at fixed $\lambda$. However, the exponential 
suppression in $\lambda$ is replaced by a power suppression in $1/D$.

 Next we calculate the pairing contribution to the energy 
density. In the weak coupling limit we have \cite{Schafer:1999fe}
\be
\label{E_pair}
E = \int  \frac{d^{D-1}p}{(2\pi)^{D-1}}
 \left\{-\frac{\Delta^2}{2\sqrt{\epsilon_p^2+\Delta^2}}
+\sqrt{\epsilon_p^2+\Delta^2}-\epsilon_p \right\}.
\ee
The integrals can be calculated in the same fashion as the 
integral that appears in the gap equation. We find
\bea 
 E &=& -\frac{\Omega_{D-1}}{(2\pi)^{D-1}}
   E_Fk_F^{D-1}\frac{\pi}{2}  
   \left(1+x^2\right)^{\alpha/2}
 \left\{ \left[\frac{1}{\alpha+2}-\frac{\alpha x^2}{2\alpha+4}\right]
   P_\alpha \left( - \frac{1}{\sqrt{1+x^2}}\right) \right. 
  \nonumber  \\
 & & \hspace{4cm}\mbox{}\left.
 + \frac{\sqrt{1+x^2} }{\alpha+2}
  P_{\alpha+1} \left( - \frac{1}{\sqrt{1+x^2}} \right)
  \right\}.
\eea
In the limit $x\to 0$ the logarithmic singularities in the 
two terms in the curly brackets cancel and the energy per 
particle is proportional to $x^2$. We find
\be
\frac{E}{A} = -\frac{D-1}{4} E_F \left(\frac{\Delta}{E_F}\right)^2.
\ee
Since $\Delta/E_F=O(1/D)$ we conclude that the pairing energy 
per particle scales as $1/D$ in the large $D$ limit. This 
implies that the pairing energy is suppressed compared to 
the contribution from the ladder sum given in equ.~(\ref{E_pp_d}). 

\section{Screening in the large $D$ limit}
\label{sec_scr}

\begin{figure}
\includegraphics[width=9cm,clip=true]{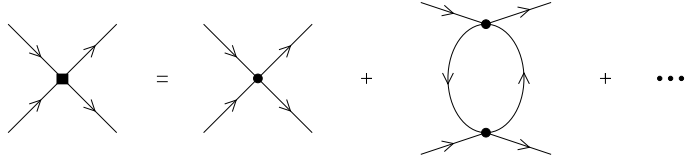}
\caption{First order screening correction to the effective
particle-particle interaction. }
\label{fig_ceff}
\end{figure}

  In $D=4$ dimensions screening of the elementary four fermion 
interaction by particle-hole pairs gives an important contribution 
to the effective interaction. It is well known, for example, that 
screening reduces the magnitude of the gap in the weak coupling
limit by a factor $(4e)^{1/3}\sim 2.2$ \cite{Gorkov:1961}. On the 
other hand, if the $1/D$ expansion corresponds to an expansion in
the number of hole lines then we expect that the screening 
correction should scale as $1/D$ in the large $D$ limit. The 
basic particle-hole bubble is given by 
\be 
\Pi(\nu,q) = \frac{Mk_F^{D-3}\Omega_{D-1}}{(2\pi)^{D-1}}
  f_{PH}(\nu,q) 
\ee
with 
\be 
 f_{PH}(\nu,q) = \frac{1}{c_D}\int_0^1 k^{D-2}dk
   \int_{-1}^1dx (1-x^2)^{D/2-2} 
    \frac{2\omega_{qk}}{\nu^2-\omega_{qk}^2},
\ee 
where $\omega_{qk}=qkx+q^2/2$. This integral can be evaluated
analytically in $D=4$ and reduces to equ.~(\ref{pi_ph}). The
particle-hole bubble leads to a renormalization of the effective
interaction as shown in Fig.~\ref{fig_ceff}. In the weak 
coupling limit only the interaction of two quasi-particles
near the Fermi surface is important. For $s$-wave pairing 
we can write
\be
\label{ceff}
C_{eff}= C_0 + C_0^2\frac{Mk_F^{D-3}\Omega_{D-1}}{(2\pi)^{D-1}}
 \bar{f}_{PH}
\ee
where $\bar{f}_{PH}$ is an average over the Fermi surface
\be 
\label{f_av}
\bar{f}_{PH} =  \frac{1}{c_D}\int_{-1}^1 dx\, 
 (1-x^2)^{D/2-2}f_{PH}\left(0,\sqrt{2(1-x)}\right).
\ee
\begin{figure}
\includegraphics[width=9cm,clip=true]{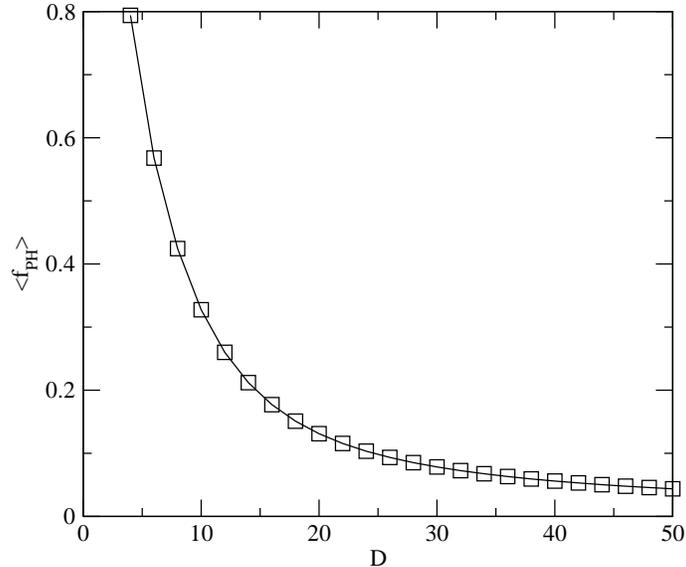}
\caption{Screening correction to the effective $s$-wave
particle-particle interaction as a function of the number 
of space-time dimensions $D$. }
\label{fig_fph}
\end{figure}
Replacing the bare interaction with the effective one 
in the BCS gap equation leads to a correction for the 
pairing gap. We find
\be 
\Delta = \frac{2e^{-\gamma}E_F}{D} 
  \exp\left(-\bar{f}_{PH}\right)
  \exp\left(-\frac{1}{D\lambda}\right).
\ee
For $D=4$ the integral in equ.~(\ref{f_av}) can also be evaluated
analytically, yielding $\bar{f}_{PH}=(2\log(2)+1)/3\simeq 0.79$. 
Numerical results for $D\geq 4$ are shown in Fig.~\ref{fig_fph}.
We observe that the screening correction vanishes as $1/D$ for 
large $D$.

\section{Conclusions}
\label{sec_sum}

 In this work we considered different many body theories 
for a system of non-relativistic fermions described by an effective 
field theory. We are interested, in particular, in a systematic 
approach to the problem of a dilute liquid of fermions in the limit 
in which the  scattering length is large compared to the inter-particle
spacing. This led us to consider the large $g$ and large $D$ 
expansions. 

\begin{figure}
\includegraphics[width=9cm,clip=true]{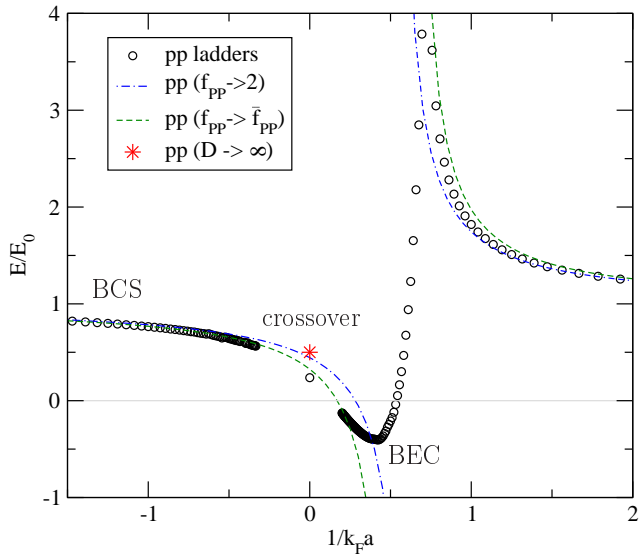}
\caption{Total energy of an interacting fermion gas in units 
of the energy of a free fermion gas as a function of $(k_Fa)^{-1}$.
The open circles show the result of a numerical calculation of 
particle-particle ladder diagrams. The dashed and dash-dotted 
curves show the two approximations discussed in Sect.~\ref{sec_pp}.
The star is the result of the $D\to\infty$ calculation in 
the unitary limit.  }
\label{fig_ring_sum}
\end{figure}

 At leading order the large $g$ expansion corresponds to summing 
all particle-hole ring diagrams with the additional approximation
that all spin factors are replaced by their large $g$ limits. The
problem is that the large $g$ limit is not suitable for studying 
the limit $|k_Fa|\to\infty$. Indeed, the naive large $g$ limit 
corresponds to taking $|k_Fa|\to 0$. This is also manifest in
the functional form of $E/A$ at leading or sub-leading order in 
$1/g$. The energy per particle in the limit $|k_Fa|\to\infty$ is 
strongly dependent on the regularization scale. The only sensible 
alternative is the Bose limit where we keep $k_Fa$ fixed and small
and take $g\to\infty$ with $\rho=gk_F^3/(6\pi^2)$ constant 
\cite{Furnstahl:2002gt,Jackson:1994}.

 From the study of the two-body system in effective field theory 
we know that if the scattering length is large then the two-body
interaction has to summed to all orders. This suggests that a 
sensible many body theory has to contain at least all particle
ladders. We have studied the particle ladder sum in effective 
field theory. If the particle-particle bubble is replaced by its
phase space average then the particle sum can be carried out 
analytically. This approximation gives $\xi=0.32$ for the universal
parameter in the equation of state \cite{Heiselberg:2000bm}. Numerically, 
we find a smaller value $\xi\simeq 0.24$. The ladder approximation 
is unreliable in the regime $(k_Fa)\sim 1$ in which deeply 
bound two-body state are important. These results are summarized
in Fig.~\ref{fig_ring_sum}. We observe that different approximations
agree in the weak coupling (BCS) limit, but give a range of
predictions for the crossover behavior.

 It is clearly desirable to construct a systematic expansion 
that contains the ladder sum at leading order. In a very interesting 
paper Steele suggested that the large $D$ expansion provides
the desired approximation scheme. In order to test this idea
we have calculated particle ladders for arbitrary $D$. We find
that if the coupling constant is scaled appropriately then 
all particle ladders are indeed of the same order in $1/D$. 
We also find that the universal parameter $\xi$ is given by
$\xi=1/2$, in surprisingly good agreement with the Green function 
Monte Carlo result $\xi=0.44$ \cite{Chang:2004sj}. We also 
verified that the contribution from the pairing energy is 
suppressed by $1/D$. 

 There are many questions regarding the $1/D$ expansion that 
remain to be addressed. We have not been able to construct a general 
method for computing $1/D$ corrections. We have also not succeeded
in showing that the $1/D$ expansion corresponds to an expansion
in the number of hole lines. Our result for the energy per particle
in the large $D$ limit differs from Steele's result $\xi=4/9$ 
because he did not compute many body diagrams for a general number 
of space-time dimensions. Instead, he performed certain kinematic 
expansions in the $D=4$ loop integrals. Even if this method was 
correct it would correspond to a partial resummation of $1/D$ 
corrections. We also believe that the expansions that are used 
in Steele's paper are not convergent. Finally, we note that there 
is an argument due to Nussinov and Nussinov \cite{Nussinov:2004} 
which indicates that for $D>4$ the ground state consists of 
non-interacting, zero energy bosons and that $\xi=0$. This argument 
may indicate that there is a subtlety with regard to the order of 
the $D\to\infty$ and $(k_Fa)\to\infty$ limits. 

Acknowledgments: This work was supported by US Department of Energy 
grants DE-FG02-03ER41260 (T.S.) and DE-FG02-97ER41048 (S.C. and C.K.). 
We would like to thank D.~Lee, H.~Hammer and T.~Mehen for useful 
discussions. After this work was finished Schwenk and Pethick 
computed effective range corrections to the equation of state 
in the unitary limit \cite{Schwenk:2005ka}. In order to facilitate
the comparison with their results we added Fig.~\ref{fig_eff_range}
to this paper. We thank C.~Pethick and A.~Schwenk for useful
correspondence regarding their work.

\appendix
\section{Effective Range Corrections}
\label{sec_app}

 In this appendix we provide some details regarding the calculation 
of effective range corrections to the particle-particle ladder sum. 
The effective interaction is given by
\be
\langle P/2\pm k_{1}|V|P/2\pm k_{2}\rangle
  =C_{0}+\frac{C_{2}}{2}(k_{1}^{2}+k_{2}^{2})
  =C_{0}+C_{2}^{L}\cdot k_{1}^{2}+C_{2}^{R}\cdot k_{2}^{2},
\ee
where $C_{2}^{L}=C_{2}^{R}=\frac{C_{2}}{2}$. We consider the 
following amplitudes
\begin{figure}
\includegraphics[width=12cm,clip=true]{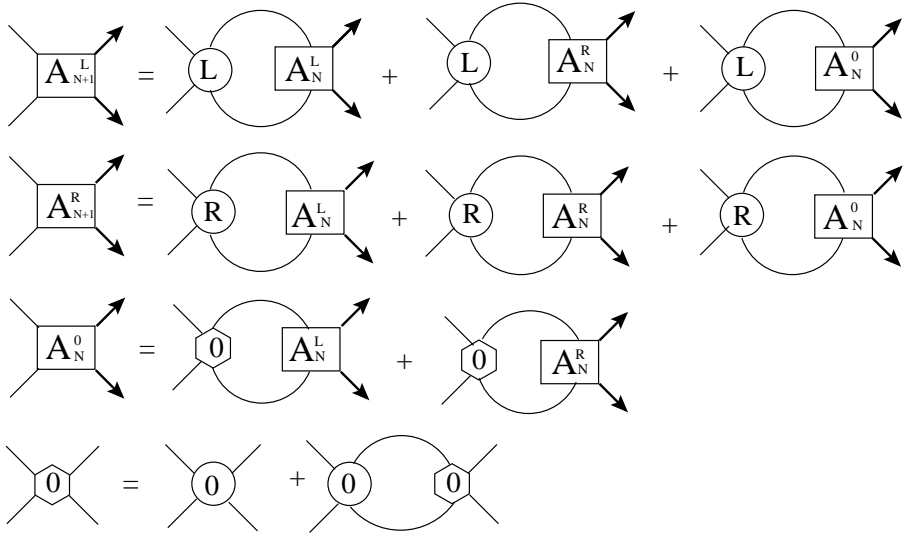}
\caption{The recursion relations between $A_{N}^{L}$,$A_{N}^{R}$ and 
$A_{N}^{0}$.}
\label{fig_rec}
\end{figure}
\begin{itemize}
\item $X_{N}^{R}$($X_{N}^{L}$): The sum of all amplitudes ${\cal T}
      (P/2\pm k,P/2\pm k)$ starting with a $R$($L$) vertex and 
      containing $N$ $C_2$ ($L$ or $R$) vertices.
\item $X_{N}^{0}$: The sum of all amplitudes starting with a $C_{0}$
      vertex and containing $N$ $C_2$ ($L$ or $R$) vertices. 
\end{itemize}
In this appendix we evaluate $\sum_{N=1}^{N=\infty}(X_{N}^{L}+X_{N}^{R}
+X_{N}^{0})$. Our strategy is to derive a set of recursion relations 
between the amplitudes and then use these relations to compute the sum. 
We define
\be
X_{N}^{L}=k^{2}\cdot A_{N}^{L},\hspace{0.5cm}
X_{N}^{R}=A_{N}^{R}, \hspace{0.5cm}
X_{N}^{0}=A_{N}^{0}.
\ee
The only difference between $X_{N}$ and $A_{N}$ is that $A_{N}$ does 
not include the contribution from the initial pair momentum $k$. 
The amplitudes $A_{N}$ satisfy (see Fig.~\ref{fig_rec})
\bea
A_{N+1}^{L}&=&\frac{C_{2}}{2}I_{2}A_{N,L}+
  \frac{C_{2}}{2}I_{0}A_{N}^{R}+\frac{C_{2}}{2}I_{0}A_{N}^{0},
  \nonumber \\
A_{N+1}^{R}&=&\frac{C_{2}}{2}I_{4}A_{N,L}+\frac{C_{2}}{2}I_{2}A_{N}^{R}
  +\frac{C_{2}}{2}I_{2}A_{N}^{0},  \\
A_{N}^{0}&=&\frac{C_0}{1-C_{0}I_{0}}[I_{2}A_{N}^{L}+I_{0}A_{N}^{R}].
  \nonumber 
\eea
Since $A_{N}^{0}$ can be expressed as the combination of $A_{N}^{L}$ and
$A_{N}^{R}$ one can simplify the above recursion relations and obtain
\bea
(1-C_{0}I_{0})A_{N+1}^{L}&=&\frac{C_{2}}{2}
 \left[I_{2}A_{N}^{L}+I_{0}A_{N}^{R}\right],
 \nonumber \\
(1-C_{0}I_{0})A_{N+1}^{R}&=&\frac{C_{2}}{2}
  \left[(I_{4}-C_{0}I_{0}I_{4}+C_{0}I_{2}^{2})
  A_{N}^{L}+I_{2}A_{N}^{R}\right].
\label{recur}
\eea
The crucial point is that the $lhs$ of equ.~(\ref{recur}) starts 
from $A_{2}$ while the $rhs$ starts from $A_{1}$. Therefore one 
obtains 
\bea
2\left(1-C_{0}I_{0}\right)
  \left( \sum_{N=1}^{\infty}A_{N}^{L}-A_{1}^{L} \right)
    &=&C_{2}I_{2}\sum_{N=1}^{\infty}
    A_{N}^{L}+C_{2}I_{0}\sum_{N=1}^{\infty}A_{N}^{R},\nonumber \\
2\left(1-C_{0}I_{0}\right)
  \left(\sum_{N=1}^{\infty}A_{N}^{R}-A_{1}^{R}\right)
    &=&C_{2}\left(I_{4}-C_{0}I_{0}I_{4}+C_{0}I_{2}^{2}\right)
    \sum_{N=1}^{\infty}A_{N}^{R}+C_{2}I_{2}\sum_{N=1}^{\infty}A_{N}^{R},
\eea
where $A_{1}^{L}$ and $A_{1}^{R}$ are given by
\bea
A_{1}^{L}&=&\frac{C_{2}}{2}+\frac{C_{2}}{2}I_{0}\cdot\frac{C_{0}}{1-C_{0}I_{0}}
  =\frac{C_{2}}{2(1-C_{0}I_{0})},\,\,\,\nonumber \\
A_{1}^{R}&=&\frac{C_{2}}{2}k^2+\frac{C_{2}}{2}I_{2}
  \cdot\frac{C_{0}}{1-C_{0}I_{0}}=
  \frac{C_{2}}{2}\left[\frac{k^2+C_{0}(I_{2}-k^{2}I_{0})}{1-C_{0}I_{0}}\right].
\eea
We define the sums over $A_{N}^{L}$, $A_{N}^{R}$ and $A_{N}^{0}$ as
\be
\sum_{N=1}^{\infty}A_{N}^{R}=Y_{R},\hspace{0.5cm}
\sum_{N=1}^{\infty}A_{N}^{L}=Y_{L},\hspace{0.5cm}
\sum_{N=1}^{\infty}A_{N}^{0}=Y_{0}.
\ee
After some algebra one obtains
\bea
Y_{L}&=&\frac{C_{2}}{2}\cdot
  \frac{(1-C_{0}I_{0})(1-C_{2}(I_{2}-k^{2}I_{0})/2)}
  {(1-C_{0}I_{0}-C_{2}I_{2}/2)^{2}-C_{2}^{2}I_{0}(C_{0}(-I_{0}I_{4}+I_{2}^{2})
  +I_{4})/4}, \nonumber \\
Y_{R}&=&\frac{C_{2}}{2}\cdot
  \frac{(1-C_{0}I_{0}-C_{2}I_{2}/2)(k^2+C_{0}(I_{2}-k^2I_{0}))
  +C_{2}(C_{0}(-I_{0}I_{4}+I_{2}^{2})
  +I_{4})/2}
  {(1-C_{0}I_{0}-C_{2}I_{2}/2)^{2}-C_{2}^{2}I_{0}(C_{0}(-I_{0}I_{4}+I_{2}^{2})
  +I_{4})/4}.
\eea
It is now straightforward to evaluate the sums over $X_{N}$ 
\bea
 \lefteqn{\sum_{N=1}^{N=\infty} \left(X_{N}^{L}+X_{N}^{R}+X_{N}^{0}\right)
    =  k^{2}\cdot Y_{L}+Y_{R}+Y_{0}} \nonumber \\
   &=&\frac{1}{1-C_{0}I_{0}}Y_{R}+\left(k^2+\frac{C_{0}I_{2}}{1-C_{0}I_{0}}
      \right)Y_{L} \nonumber \\
   &=&\frac{C_{2}}{2(1-C_{0}I_{0})}\cdot
    \frac{(1-C_{0}I_{0}-C_{2}I_{2}/2)(k^2+C_{0}(I_{2}-k^2I_{0}))
    +C_{2}(C_{0}(-I_{0}I_{4}+I_{2}^{2})
    +I_{4})/2}
    {(1-C_{0}I_{0}-C_{2}I_{2}/2)^{2}-C_{2}^{2}I_{0}
    (C_{0}(-I_{0}I_{4}+I_{2}^{2})+I_{4})/4}. \nonumber \\
   & & \mbox{}+\frac{C_{2}}{2}\cdot
   \frac{((k^2+C_{0}(I_{2}-k^2I_{0}))(1-C_{2}(I_{2}-k^2I_{0})/2)}
        {(1-C_{0}I_{0}-C_{2}I_{2}/2)^{2}-C_{2}^{2}I_{0}(C_{0}
         (-I_{0}I_{4}+I_{2}^{2})+I_{4})/4}.
\label{eq:result}
\eea
Equ.~(\ref{eq:result}) can be used to derive equ.~(\ref{E_pp_C2}). 


\end{document}